\documentclass[aps,pre]{revtex4-1}
\usepackage{amsmath,amssymb}
\usepackage{graphicx}

\usepackage[abs]{overpic}

\usepackage[hypertex]{hyperref} 
\include{epsf}

\def\(({\left(}
\def\)){\right)}                       
\def\[[{\left[}
\def\]]{\right]}

\newcommand{\<}{\langle}
\renewcommand{\>}{\rangle}
\newcommand{\be}{\begin{equation}}
\newcommand{\ee}{\end{equation}}
\newcommand{\bea}{\begin{eqnarray}}
\newcommand{\eea}{\end{eqnarray}}

\renewcommand{\b}{\beta}

\newcommand{\de}{\delta}
\newcommand{\ep}{\varepsilon}

\newcommand{\ex}{\text{e}}

\begin{document}
\title{Exact mean field inference in asymmetric kinetic Ising systems}

\author{M. M\'ezard and J. Sakellariou}
\affiliation{
Laboratoire de Physique Th\'eorique et Mod\`eles Statistiques,
CNRS and Universit\'e Paris-Sud, B\^{a}t 100, 91405 Orsay Cedex, France}

\begin{abstract}
We develop an elementary mean field approach for fully asymmetric
kinetic Ising models, which can be applied to a single instance of the
problem. In the case of the asymmetric SK model this method gives the
exact values of the local magnetizations and the exact relation
between equal-time and time-delayed correlations. It can also be
used to solve efficiently the inverse problem, i.e. determine the
couplings and local fields from a set of patterns, also in cases where the fields and
couplings are time-dependent. This approach generalizes some recent
attempts to solve this dynamical inference problem, which were valid
in the limit of weak coupling. It provides the exact solution to the
problem also in strongly coupled problems. This mean field inference
can also be used as an efficient approximate method to infer the couplings and
fields in problems which are not infinite range, for instance in
diluted asymmetric spin glasses.

\end{abstract}

\date{\today}

\maketitle

Inference problems are as old as scientific modelling: given data, how
can
we reconstruct a model which accounts for it, and find the parameters
of the model? This is particularly difficult when data is obtained
from networks of many interacting components.
The fast development of high-throughput technologies in various fields
of biology, ranging from gene regulation to protein interaction and
neural activity,  is
generating a lot of data, which is challenging our ability to infer
the structure and parameters of the underlying networks.

This `network reconstruction' problem is typically an inverse problem
which has motivated a lot of activity  in machine learning
and  in statistical
physics\cite{AckleyHinton,Hinton,KappenSpanjers,KappenRodriguez,Tanaka,SessakMonasson,MezardMora,Marinari,HuangSuscProp,HuangMF,Wainwright,AurellOR,HertzRTTAZ,RoudiHertzPRL,RoudiHertzLong,RoudiAH,ZengAAM,WeigtWhite,BraunsteinPWZ,CoccoLeibMon}. 
Until recently the main efforts have been dedicated to
reconstructing equilibrium Boltzmann-Gibbs distributions. In the
so-called inverse
Ising model, one typically assumes to have data in the form of some
configurations,
which we shall call `patterns', of a
$N$-spin Ising system drawn from the Boltzmann-Gibbs distribution with an
energy function including one-body (local magnetic fields) and
two-body (exchange couplings) terms. The problem is to reconstruct the
local fields and the exchange couplings (collectively  denoted below
as `couplings') from the data. 
This problem has been actively studied in recent years, in particular in the context of
neural network inference based on multielectrode recordings in
retinas \cite{SchneidBSB,ShlensFGG,CoccoLeibMon}.
The standard
solution of this problem, known as the Boltzmann machine, computes,
for some given couplings, the local magnetizations and the two-spin
correlations, and compares them to the empirical estimates of
magnetizations and correlations found from the patterns\cite{AckleyHinton,Hinton}. The couplings
are then iteratively adjusted in order to decrease the distance
between the empirical magnetizations/correlations and the ones
computed from the model. A Bayesian formulation shows that the problem
of finding the couplings is actually convex, so that this iterative
procedure is guaranteed to converge to the correct couplings, provided
that the number of patterns is large enough to allow for a good
estimate of correlations. The drawback of this method is that the
reliable computation of the magnetizations/correlations, given the couplings,
which is done using a Monte Carlo procedure, is extremely time
consuming. Therefore this exact approach is limited to systems with
a small number of spins. Most of the recent works on this issue have
developed  approximate methods to infer the
couplings. Among the most studied approaches are the naive mean
field method \cite{KappenRodriguez,Tanaka,HertzRTTAZ}, the TAP approach \cite{KappenSpanjers,RoudiHertzPRL,ZengAAM}, a method based on
a small magnetization expansion \cite{SessakMonasson}, and a message-passing
method called susceptibility propagation\cite{MezardMora,Marinari,HuangSuscProp}. Another approach
which has been developed is that of linear relaxation of the inference problem\cite{Wainwright}.
The inverse-Potts problem is a version
of this same problem, with variables taking $q$ possible states. The
case $q=20$ is relevant for inferring interaction in protein pairs
from data on co-evolution of these pairs, and its solution by
susceptibility propagation has given an accurate prediction  of
inter-protein residue contacts\cite{WeigtWhite}.
Another case which has received some attention is the problem of
reconstruction in Boolean networks (see e.g.\cite{BraunsteinPWZ}
and references therein).

However, in many applications to
biological systems, in particular the ones concerning neural activity
and gene expression network, the assumption that the patterns are
generated by an underlying equilibrium Boltzmann-Gibbs measure is not
well founded. Couplings are typically asymmetric, and they may vary in
time, so there is no equilibrium measure.
This has prompted the recent study of inference in purely kinetic
models without an equilibrium measure
\cite{CoccoLeibMon,HertzRTTAZ,RoudiHertzPRL,ZengAAM}. A benchmark on this dynamic
inference problem is the inverse asymmetric kinetic Ising
model. The framework is the same as the equilibrium one: one tries to
infer the parameters of the dynamical evolution equation of an
Ising spin systems, given a set
of patterns generated by this evolution. The recent works \cite{HertzRTTAZ,RoudiHertzPRL,RoudiHertzLong,ZengAAM} have studied the performance of
two mean-field methods on this problem, the naive mean field (nMF) and
a weak-coupling expansion which they denote as TAP method. They have shown that, in the case of the fully
asymmetric infinite range spin glass problem, the inference
problem can be solved by these methods in the case where the spins
are weakly coupled. In the present work we present a (non-naive!) mean field
approach which solves the problem at all values of the couplings (and
reduces to their TAP approach at weak coupling).

\begin{figure}
    \includegraphics[width=0.95\linewidth]{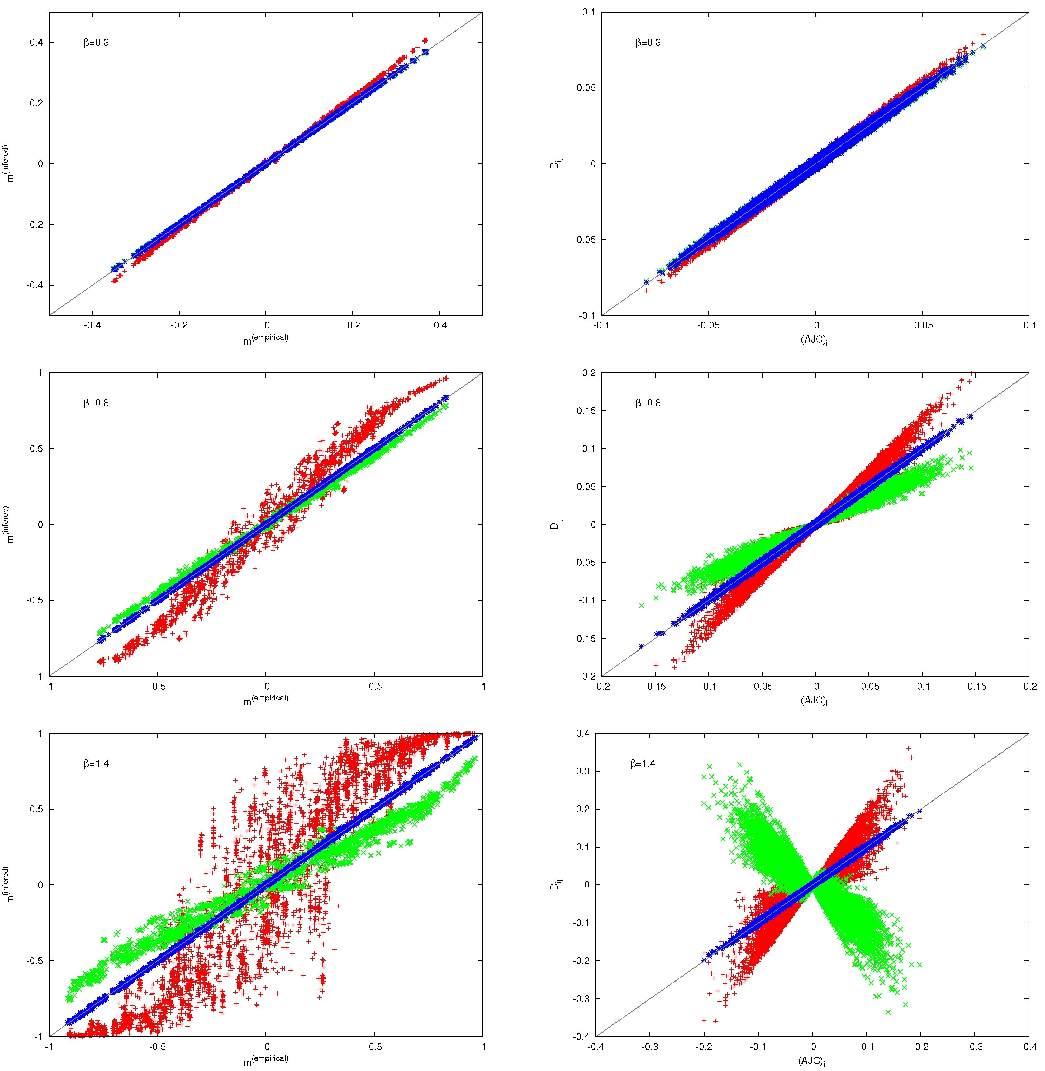}
  \caption{Magnetizations (left column) and correlations (right column) obtained by MF (blue), `TAP' (green) and
    nMF (red). One $N=200$ spin model is simulated $500000$ times for $31$ time steps. 
    The three plots in each column correspond to inverse temperature $\b=.3,\; .8$ and $1.4$ (from top to bottom).
   In the left column, the magnetizations predicted by each method for all time steps are plotted versus the experimental ones
    found by monte carlo simulation.
    For the plots of the right column, the correlation matrices $C$ and $D$ are obtained at $t=30$. The scatter plot
    shows for each pair $ij$, the value of $D_{ij}$ in ordinate, and
    the value of $ (A J C)_{ij}$ in abscissa. The three methods differ in
    their predictions for $A$. At high temperature, $\beta=.3$, all methods are
    good for both the magnetizations and correlations; the MF and `TAP' methods nearly coincide and are slightly
    better than nMF. At larger and larger $\beta$, the `TAP'
    correction to naive mean field overshoots, and only the MF results
    is correct. The data supports the statement that MF is exact at
    all temperatures, while nMF and `TAP' are only high temperature approximations.
  }  
  \label{fig:PbDirect}
\end{figure}

The kinetic Ising model which we shall study is the same as the one of
\cite{RoudiHertzPRL}.
$N$ Ising spins $s_i$ evolve in discrete time, with a synchronous
parallel dynamics. Given the configuration of spins at time $t-1$,
$s(t-1)=\{s_1(t-1),\dots,s_N(t-1)\}$, the spins $s_i(t)$ are
independent random variables drawn from the distribution:
\be
\label{Distr1Spin1Time}
P(s(t)|s(t-1))=\prod_{i=1}^N \frac{1}{2 \cosh(\beta h_i(t))}e^{\beta s_i(t) h_i(t)}
\ee
where
\be
\label{Field1Spin1Time}
h_i(t)=H_i(t-1)+\sum_j J_{ij}(t-1) s_j(t-1)
\ee
Note that both the local external fields $H_i(t)$ and the exchange couplings
$J_{ij}(t)$ may depend on time.
Here we are interested in a fully asymmetric model. We generate the
$J_{ij}$ by an asymmetric version \cite{HertzGS,Parisi,Derrida} of the
infinite -range Sherrington-Kirkpatrick spin glass model \cite{SK}, in which for each directed edge $(ij)$ the coupling is a gaussian
random variable with variance $1/N$.
Notice that $J_{ij}$ and $J_{ji}$ are independent random variables. We
do not include self-interactions (we take $J_{ii}=0$), although
this could be done without changing the results.
As initial conditions we take $s_i(t=0)=\pm 1$ with probability
$1/2$.  Our method also applies to the case of asynchronous dynamics,
studied in \cite{ZengAAM} with the TAP approach, but to keep the
presentation simple we shall study only  the case of the
synchronous parallel dynamics in
this letter.


We first derive the mean-field equations for the magnetizations
$m_i(t)=\langle s_i(t)\rangle$. Because the couplings are
asymmetric, $\sum_j J_{ij}J_{ji}= O(1/\sqrt{N})$, therefore the
Onsager reaction term is not present in
this problem. This makes the derivation of our equations, which we shall denote
just 'mean-field' equations, particularly easy. The approximate equations 
used in \cite{RoudiHertzPRL,ZengAAM}, originally derived in
\cite{KappenRodriguez}, have been obtained by a second order expansion in the couplings. When this expansion is applied to the symmetric problem it gives back the TAP equations \cite{TAP} with their Onsager reaction term. In the present case of asymmetric coupling, it keeps the correction of order $ \sum_j J_{ij}J_{ij}$. We shall keep for these second-order-expanded equations  the name `TAP'-equations, as used by \cite{KappenRodriguez,RoudiHertzPRL,ZengAAM}.

The local field on spin $i$ due to the other
spins, $\sum_j J_{ij}(t-1) s_j(t-1) $, is the sum of a large number
of terms. Therefore it has a gaussian distribution with mean
\be
g_i(t-1)=\sum_j J_{ij}m_j(t-1)
\label{MF1}
\ee
and variance
\be
\Delta_i(t-1)=\sum_j J_{ij}^2(1-m_j(t-1)^2)
\label{MF2}
\ee
(in order to derive this last formula, one must use the fact that the
typical connected correlation $\langle s_j s_k\rangle-m_jm_k$ is
typically of order $1/\sqrt{N}$; this will be checked
self-consistently below). Using this property and the definition of
the dynamics (\ref{Distr1Spin1Time}),  one obtains the
magnetization of spin $i$ at time $t$:
\be
\label{MF3}
m_i(t)= 
\int D x \; \tanh \left[\b \left( 
H_i(t-1)+ g_i(t-1) + x\sqrt{\Delta_i(t-1)} \right)  \right] \ ,
\ee
where $Dx= \frac{dx}{\sqrt{2 \pi}} \ex^{-\frac{x^2}{2}}$ is the
probability density for a Gaussian variable $x$ with zero mean and
variance unity. 

Equations (\ref{MF1},\ref{MF2},\ref{MF3}) are our mean field (MF) equations for this
problem, valid on a given instance. Similar
dynamical equations have been obtained in the study of the sample-averaged order parameter in asymmetric neural
networks\cite{DerGarZip,GutMez} and spin glasses\cite{Derrida}. They can be iterated starting from some initial condition (in
our case $m_i(0)=0$) in order to get all  the magnetizations $m_i(t)$ at
any time. They rely only on the central limit theorem and they are
exact in the large $N$ limit, for any set of couplings and
external fields, even if they are time-dependent.
These differ from the `TAP' equations of
\cite{RoudiHertzPRL,ZengAAM,KappenRodriguez,KappenSpanjers} which can be written in our notation:
\be
\label{TAProudi}
m_i(t) = \tanh \left[ \b  H_i(t-1)+ \b g_i(t-1)-m_i(t) \b^2 \Delta_i (t-1)
\right]\ , 
\ee
and from the naive mean field (nMF) equations:
\be
\label{TAPnaive}
m_i(t) = \tanh \left[\b \left(  H_i(t-1)+g_i(t-1)
\right)\right]\ .
\ee
The nMF equations and the `TAP' equations actually give
the same result as our exact MF equations, when expanded in powers of
$\Delta_i$, respectively to order $\Delta_i^0$ and $\Delta_i^1$, but they differ at order
$\Delta_i^2$. The fact that `TAP' equations agree with the exact MF to
second order in a weak coupling expansion is consistent with their
derivation through second order Plefka-type expansion\cite{KappenSpanjers}.
The correctness of the MF equations (\ref{MF3},\ref{MF1},\ref{MF2})
can be easily checked numerically as shown in the left panels of
Fig.\ref{fig:PbDirect}.

We now turn to the computation of correlations. We shall establish the mean field
relation between the time-delayed and the equal-time
correlation matrices:
\bea
D_{ij}(t)& \equiv& \<\de s_i(t+1) \de s_j(t) \> \\
C_{ij}(t) &\equiv& \<\de s_i(t) \de s_j(t) \>\ ,
\label{DCdef}
\eea
where
we define $\de s_i(t)$ as the fluctuation of the magnetization:
$
\de s_i(t)=s_i(t)-\langle s_i(t)\rangle\ .$

We start by writing 
$\sum_j J_{ij}(t) s_j(t) = g_i(t) + \de g_i(t)$, where ${\de g_i}(t)$ is gaussian distributed with mean $0$ and 
variance $\Delta_i(t)$. Now, by definition of $D_{ij}$ we have

\be
\label{DijDeff}
  D_{ij}(t) = \< s_j(t) \tanh\left[ \b \left(H_i(t) + g_i(t) + {\de
        g_i}(t) \right) \right] \> 
  -\< s_j(t)\> \< \tanh\left[ \b \left(H_i(t) + g_i(t) + {\de g_i}(t)  \right) \right] \>
\ee 
Hereafter in order to keep notations simple in the derivation of the
relation between $D(t)$ and $C(t)$ we work at a fixed time $t$ and we thus drop the explicit
time indices: all time indices in this derivation are equal to $t$
(e.g. $J_{ij}=J_{ij}(t)$, $\de s_i=\de s_i(t)$, $g_i=g_i(t)$ etc.) We get:
\be
\label{JD}
\begin{split}
  \sum_k J_{jk} D_{ik} &= \<\left( g_j + \de g_j \right) \tanh\left[ \beta \left(H_i + g_i + \de g_i \right) \right]  \> -
  g_j  \<\tanh\left[ \beta \left(H_i + g_i + \de g_i \right) \right] \> \\
  &= \< \de g_j  \tanh\left[ \beta \left(H_i + g_i + \de g_i \right) \right] \>
\end{split}
\ee

In order to evaluate the average we need the joint distribution of
$\de g_i$ and $\de g_j$. 
The crucial point to keep in mind is that, as the couplings are of order $1/\sqrt{N}$,
each matrix element of $C$ and $D$ is also small, of order
$1/\sqrt{N}$. Their covariance is therefore small:
\bea
\label{DeltasCovar}
  \langle \de g_i  \de g_j  \rangle&=& \< \sum_k J_{ik} \left( s_k  - \<s_k \> \right) \sum_l J_{jl} \left( s_l  - \<s_l \> \right) \>  \\ 
  &=& \sum_{k,l} J_{ik} J_{jl} C_{kl} = \left( J C J^T \right)_{ij}
  \equiv \ep\ ,
\eea
where $\ep$ is typically of order $1/\sqrt{N}$.
So the joint distribution of $x=\de g_i$ and $y=\de g_j$ takes
the form, in the large $N$ limit (omitting terms of order $\ep^2$):
\be
\label{Jointxy}
  P( x,y )=\frac{1}{2\pi\sqrt{\Delta_i\Delta_j}}\exp\left( -\frac{x^2}{2 \Delta_i} -\frac{y^2}{2 \Delta_j} +\ep \frac{x y}{\Delta_i\Delta_j} \right)
\ee

Using the small $\ep$ expansion of eq. \eqref{Jointxy} we can rewrite eq. \eqref{JD} as
\bea
\label{JDGauss}
  \sum_k J_{jk} D_{ik} &=& \frac{\ep}{\Delta_i\Delta_j} \int \frac{dx}{\sqrt{2 \pi \Delta_i}} \frac{dy}{\sqrt{2 \pi \Delta_j}}
  \ex^{-\frac{x^2}{2 \Delta_i} -\frac{y^2}{2 \Delta_j}} x y^2  
  \tanh\left[ \b \left( H_i + g_i + x \right)\right] \\
  &=& \ep \b \int \frac{dx}{\sqrt{2 \pi \Delta_i}} \exp^{-\frac{x^2}{2 \Delta_i}} 
  \left( 1 - \tanh^2\left[  \b \left(H_i + g_i + x \right) \right]\right)\ .
\eea

Combining eq. \eqref{DeltasCovar} and eq. \eqref{JDGauss} we get:
\be
\label{DJ^T}
\left( D J^T \right)_{ij} = \left( J C J^T \right)_{ij} \b \int \frac{dx}{\sqrt{2 \pi \Delta_i}} \ex^{-\frac{x^2}{2 \Delta_i}} 
  \left( 1 - \tanh^2 \b \left(H_i + g_i + x \right) \right)\ ,
\ee
which gives the explicit mean-field relation between $C$ and $D$. Putting back
the time indices, we obtain the final result in matrix form:
\be
\label{DeqAJC}
D(t) = A(t)\; J(t)\; C(t) \ ,
\ee
where $A(t)$ is a diagonal matrix: $A_{ij}(t)=a_i(t)\delta_{ij}$, with:
\be
\label{A}
a_i(t)=\b \int Dx  \; 
  \left[ 1 - \tanh^2 \b \left(H_i (t) + g_i (t) + x \sqrt{\Delta_i (t)}\right) \right] \ .
\ee

The final result (\ref{DeqAJC}) takes exactly the same form as the one
found with the naive mean field equation and with the `TAP'
approach. The predictions of all three
methods, nMF, `TAP' and our MF method is always $D(t)=A(t)\;  J(t)\;  C(t)$, with a
diagonal matrix $A(t)$ which differs in each case. 
As shown in \cite{RoudiHertzPRL}, the nMF approximation gives:
\be
\label{AnMF}
a_i^{nMF}(t)=\b\left[1-m_i(t+1)^2\right]\ ,
\ee
the `TAP' approximation gives:
\be
\label{AnTAP}
a_i^{TAP}(t+1)=\b\left[1-m_i(t+1)^2\right]
\left[1-(1-m_i(t+1)^2) \b^2 \sum_k J_{ik}^2(1-m_k(t)^2)\right]
\ee
and our mean field prediction is the one given in (\ref{A}).

We claim that, as in the case of the magnetizations, our mean field
equations connecting $D$ to $C$ are exact in the asymmetric SK model, in the large $N$
limit. This statement
can be checked numerically by comparing $(AJC)_{ij}$ with the experimental 
values of $D_{ij}$ found by monte carlo simulations, as shown in Fig.\ref{fig:PbDirect}.

\begin{figure}
  \begin{minipage}{.45\textwidth}
    \includegraphics[width=0.95\linewidth]{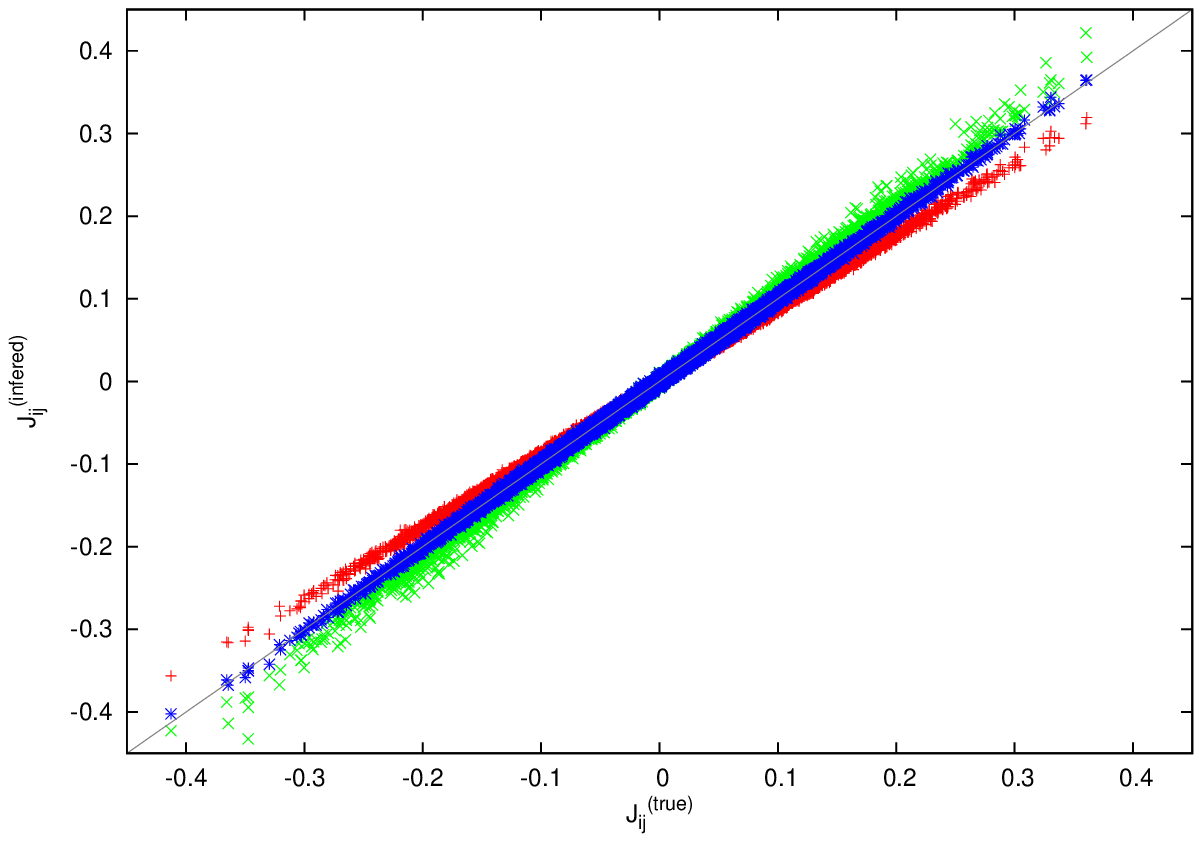}
  \end{minipage}
  \begin{minipage}{.45\textwidth}
    \includegraphics[width=0.95\linewidth]{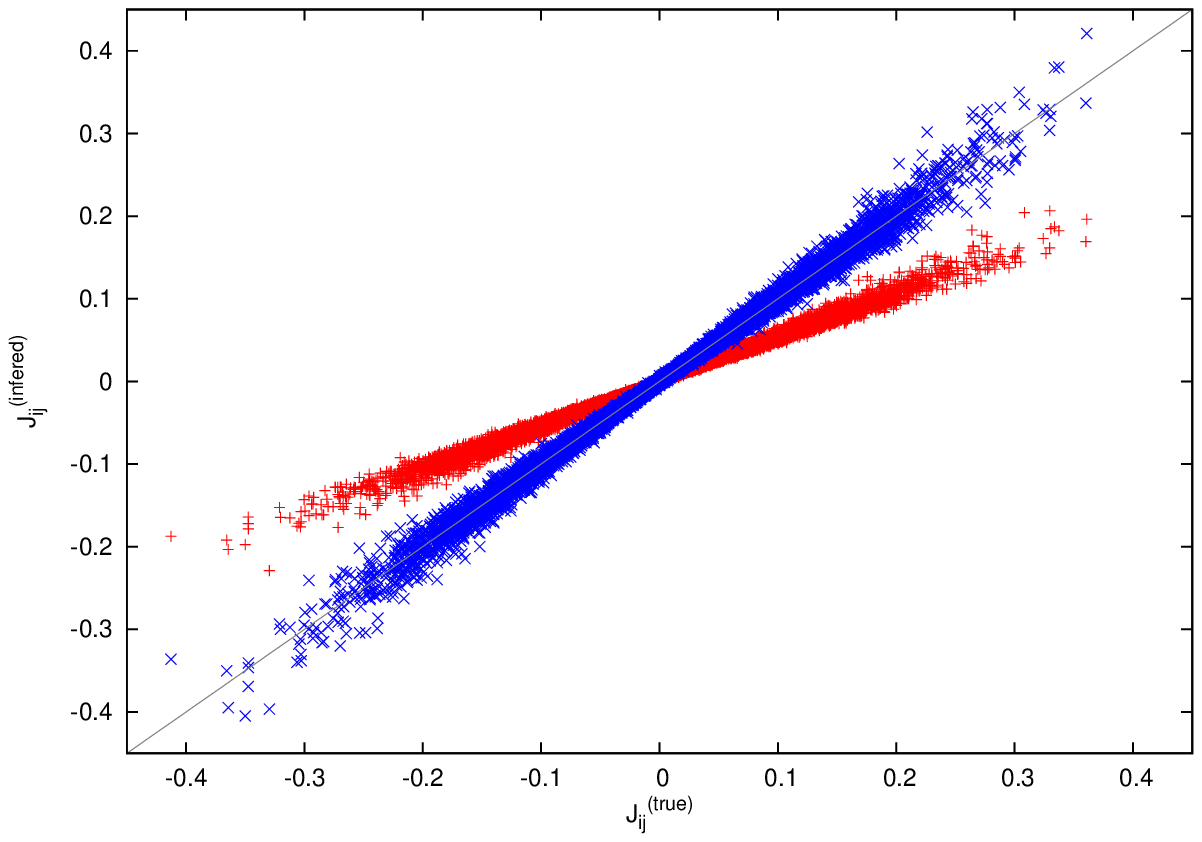}
  \end{minipage}
  \caption{Left: The infered couplings found by MF (blue), `TAP'
    (green) and nMF (red) plotted versus the real ones for a $N=100$
    model, given $P=1000000$ patterns generated at inverse temperature
    $\b=0.4$. Right: The same for $\b=1.4$ (MF (blue) and nMF (red),
    `TAP' is not shown as it fails at this high $\b$)}
  \label{fig:jjscat}
\end{figure}

\begin{figure}
  \begin{minipage}{.45\textwidth}
    \includegraphics[width=0.95\linewidth]{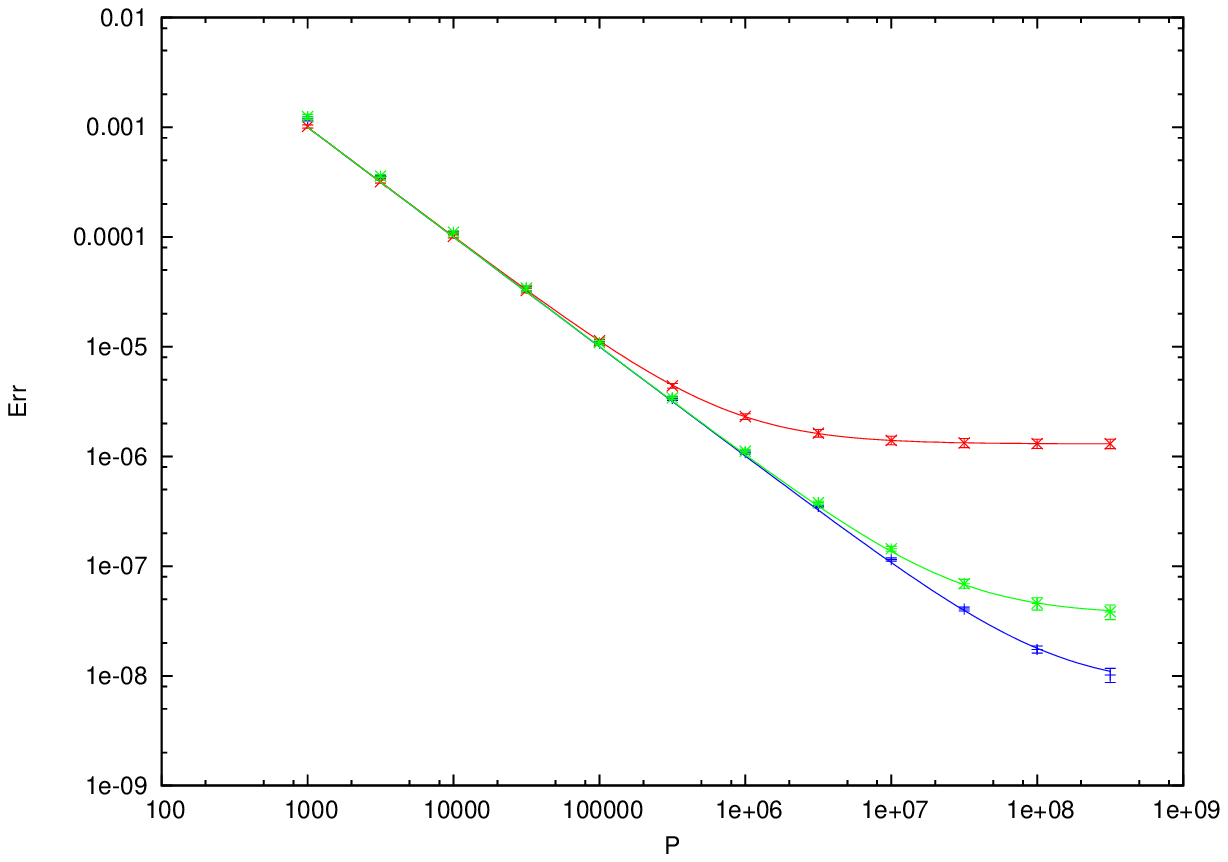}
  \end{minipage}
  \begin{minipage}{.45\textwidth}
    \includegraphics[width=0.95\linewidth]{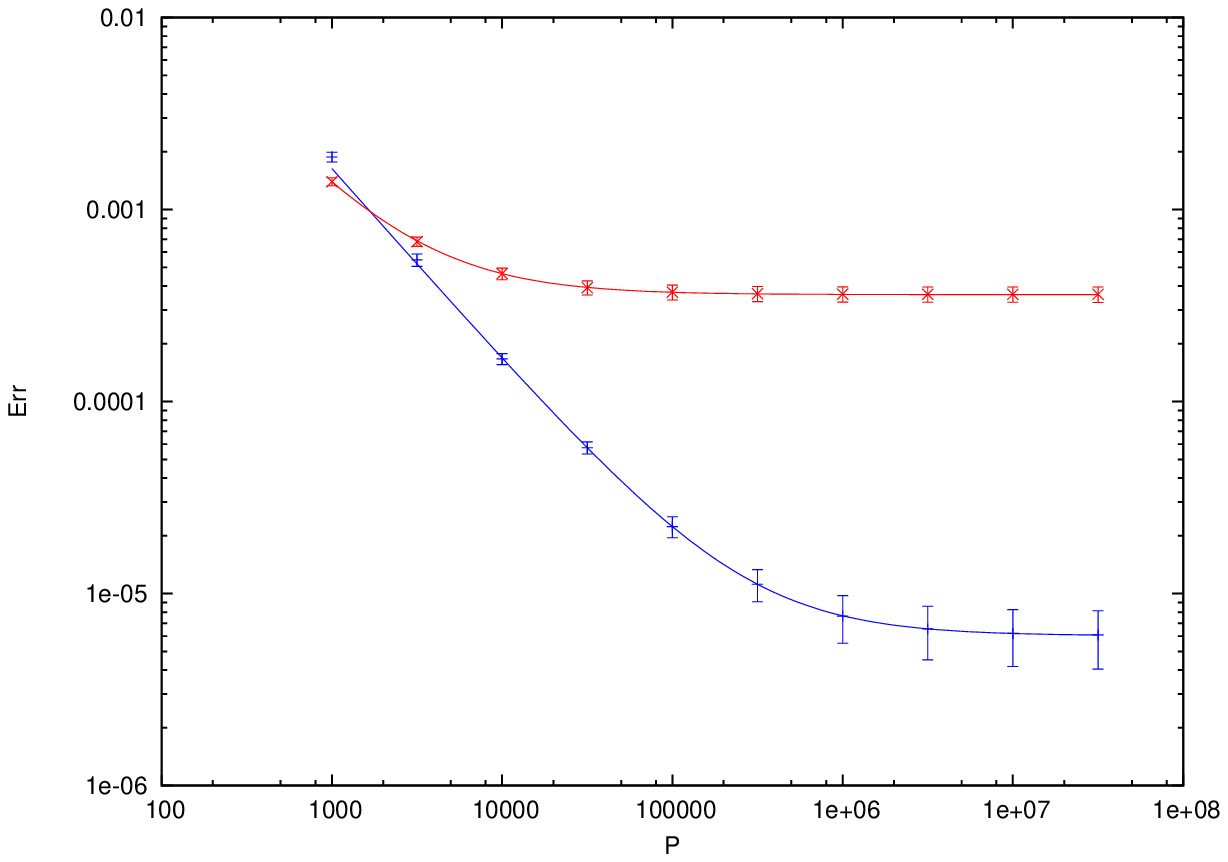}
  \end{minipage}
  \caption{Mean square error of the infered couplings
    $\overline{(J_{ij}^{\text{infered}} - J_{ij}^{\text{real}})^2}$
    obtained by MF inference (blue), `TAP' (green) and nMF
    (red) versus the number of patterns used to estimate the
    correlations, for a system of size $N=40$, where the patterns were
    generated from a gaussian distribution with a root-mean-square $\b=0.2$ (left) and $\b=0.6$
    (right). The curves are averages performed over $20$ realizations
    of the couplings and fields. Notice that the `TAP' method is absent
    in the right figure because it fails to provide results at strong
    coupling.}
  \label{fig:ErrVsL}
\end{figure}

These results on the mean field relation between $C$ and $D$ can be
used for the inverse problem. Given $P$ `patterns', which are time
sequences of length $t$ generated from the distribution
(\ref{Distr1Spin1Time}), one can estimate for each
$\tau=1,\dots,t$,  the magnetizations
$m_i(\tau)$, the equal time correlations $C_{ij}(\tau)$ and the
time-delayed correlations $D_{ij}(\tau)$. The problem is to infer from
these data the values of the couplings $J_{ij}(\tau)$ and of the local
fields $H_i(\tau)$. Without loss of generality, we can use $\beta=1$ as it is absorbed in the
strength of couplings and fields that we want to infer. We shall solve this problem using the mean field
equations. 

The problems corresponding to different times and sites
decouple. So let us consider a fixed value of $i$ and $\tau$, and
infer the $J_{ij}(\tau)$ for $j=1,\dots, N$, and $H_i(\tau)$. To
lighten notation we drop the explicit  indices $\tau$ and $i$,  and we
denote $H=H_i(\tau)$, $m_j=m_j(\tau)$,
$m=m_i(\tau+1)$,
$g=g_i(\tau)$, $\Delta=\Delta_i(\tau)$, $a=a_i(\tau)$.
Following \cite{RoudiHertzPRL}, one can obtain $J$ by inverting the relation
(\ref{DeqAJC}). The first step is to  invert the empirical $C$ matrix
and compute:
\be
b_j=\sum_k D_{ik}(\tau) C^{-1}_{kj}(\tau)\ .
\label{bDeff}
\ee
 If one knows the
number $a=a_i(\tau)$ one can then infer the couplings from (\ref{DeqAJC}):
\be
J_{ij}(\tau)=b_j/a\ .
\ee
Let us now see how $a$ can be computed.
The mean field equation (\ref{MF3}) for the magnetization reads:
\be
m= \int Dx\; \tanh\left[H+g+x\sqrt{\Delta}\right]\ .
\label{eqm}
\ee
The equation (\ref{A}) for $a$ is
\be
a=\int Dx \left(1-\tanh^2\left[H+g+x\sqrt{\Delta}\right]\right)\ .
\label{eqa}
\ee
The link between $a$ and $\Delta$ is obtained from (\ref{MF2}), which
reads:
\be
\Delta=\frac{1}{a^2}\sum_j b_j^2 (1-m_j^2)=\frac{\gamma}{a^2}\ .
\label{deltanew}
\ee
To solve this system of equations, we propose the following iterative
procedure. 
Using the empirical correlations and magnetizations estimated from the patterns,
 we first compute from (\ref{bDeff}) the $\{b_j\}$, $j\in\{1,\dots,N\}$, and
$\gamma= \sum_j b_j^2 (1-m_j^2)$.

Then we use the following mapping to find $\Delta$. 

\begin{itemize}
\item Start from a given value of $\Delta$.
\item 
Using the empirical value of $m$ and the value of $\Delta$, compute
$H+g$ by inverting (\ref{eqm}). The right-hand side of this equation
is an increasing function of $H+g$ so this inversion is easy.
\item
Using $H+g$ and $\Delta$, compute $a$ using (\ref{eqa})
\item
Compute the new value of $\Delta$, called $\hat \Delta$, using (\ref{deltanew}).
\end{itemize}

It is worth pointing out that in the thermodynamic limit, $N \to \infty$, the value of $\Delta$ becomes independent of $i$.
So, if the system under consideration is large enough, the above iteration could be perfomed only once in order to reduce
computation time.

This procedure defines a mapping from $\Delta$ to $\hat\Delta=f(\Delta)$,
and we want to find a fixed point of this mapping. It turns out that
a simple iterative procedure, starting from an arbitrary $\Delta_0$
(for instance $\Delta_0=1$)
and using $\Delta_{n+1}=f(\Delta_n)$, usuallly converges. More
precisely, it can be shown that $f(0)=\frac{\gamma}{(1-m^2)^2}$ and
that the asymptotic form for the slope of $f$ for $\Delta \gg 1$ is
$f' \sim \frac{\pi}{2} \gamma \text{exp}(\hat{u}^2) \Delta_n $, where
$\hat{u}$ is such that $m=\text{erf}(\hat{u}/\sqrt{2}) $. We have found numerically that when the number of patterns is
large enough the slope verifies: $df/d\Delta \in ]0,1[$.  Therefore
the mapping converges exponentially fast to the unique fixed point.
This method therefore works when the number of patterns per spin  $P/N$ is large
enough. In the double limit $P,N \to \infty$ and $P/N$ large enough
the above procedure thus allows to get the exact result for $\Delta$;
and therefore to find the couplings $J_{ij}(\tau)=b_j/a$. Once the
couplings have been found, one can easily compute
$g=\sum_jJ_{ij}(\tau)m_j(\tau)$, and therefore get the local field $H(\tau)$. The number
of operations needed for the full inference of the couplings and
fields is dominated by the inversion of the correlation matrix
$C$, a time which is typically at most of order $N^3$. If the number
of patterns is too small, it may happen that there is no solution to
the fixed point equation $f(\Delta)=\Delta$. Then one can decide to
use $\Delta=f(0)$, which is nothing but the nMF estimate for
$a_i(\tau)$.

\begin{figure}
 \centering
 \includegraphics[scale=1]{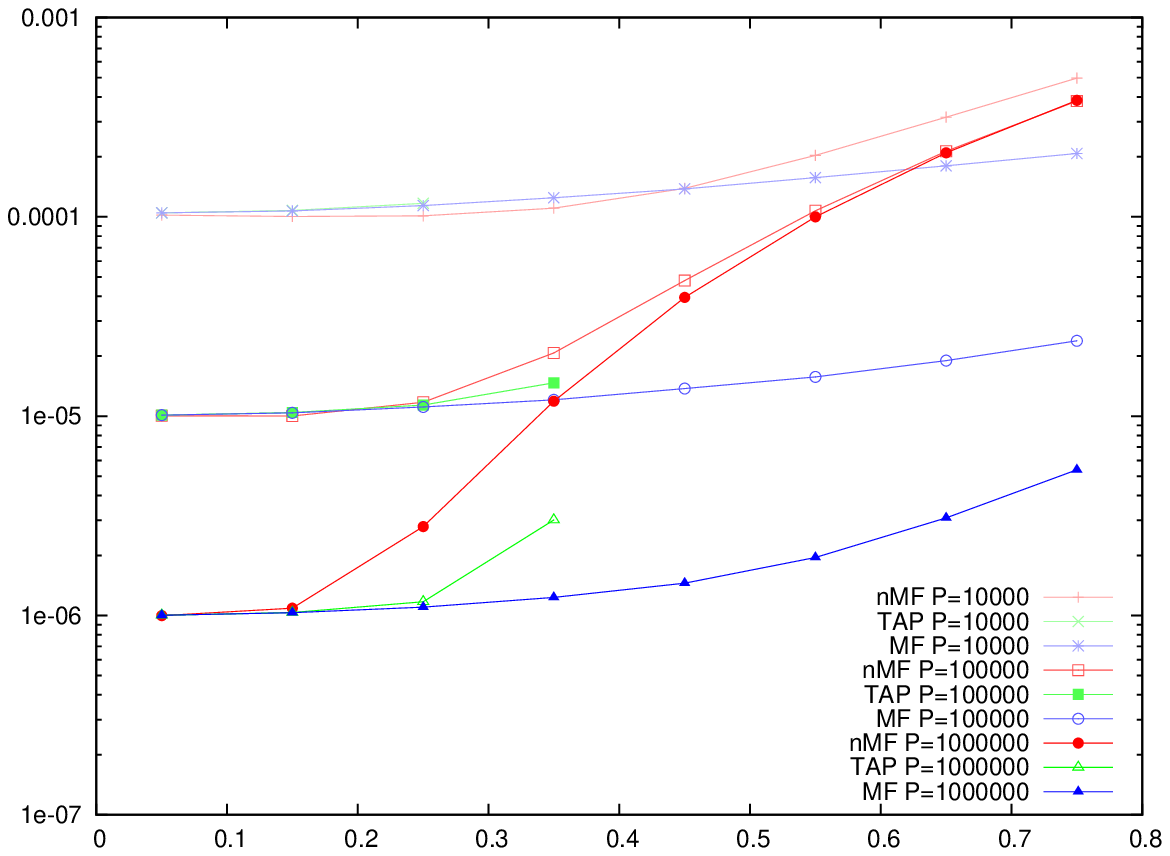}
  \caption{Mean square error of the reconstructed couplings versus
    $\beta$, averaged over $10$ systems with $100$ spins, using  the three
    inference methods nMF, `TAP' and MF, with a  number of patterns $P=10000$, $P=100000$ and $P=1000000$. 
    All three methods agree at small $\b$. The nMF error can increase by several
    orders of magnitude at large $\beta$. The `TAP' method fails to provide results above $\b \approx 0.4$.
    The MF inference method gives good results in the whole range of $\beta$.
  }
\label{fig:ErrVsBeta}
\end{figure}

We have tested our mean field inference method on the asymmetric SK
problem, where the couplings $J_{ij}$ are time-independent, gaussian distributed with
variance $\beta^2$ and the fields are time independent, uniformly
distributed on $[-\beta,\beta]$ . Fig.(\ref{fig:jjscat}) shows a
scatter plot of the result on one given instance at $\beta=.4$ and
$\beta=1.4$, and compares it to the inference method of
\cite{RoudiHertzPRL} using nMF and `TAP' (the `TAP' inference is limited
to small values of $\beta$: at large $\beta$ it fails).
Figs. \ref{fig:ErrVsL} and \ref{fig:ErrVsBeta} show a statistical
analysis of the performance of MF inference. It accurately infers the
couplings and fields even in the strong coupling regime.

\begin{figure}
    \includegraphics[width=0.5\linewidth,angle=0]{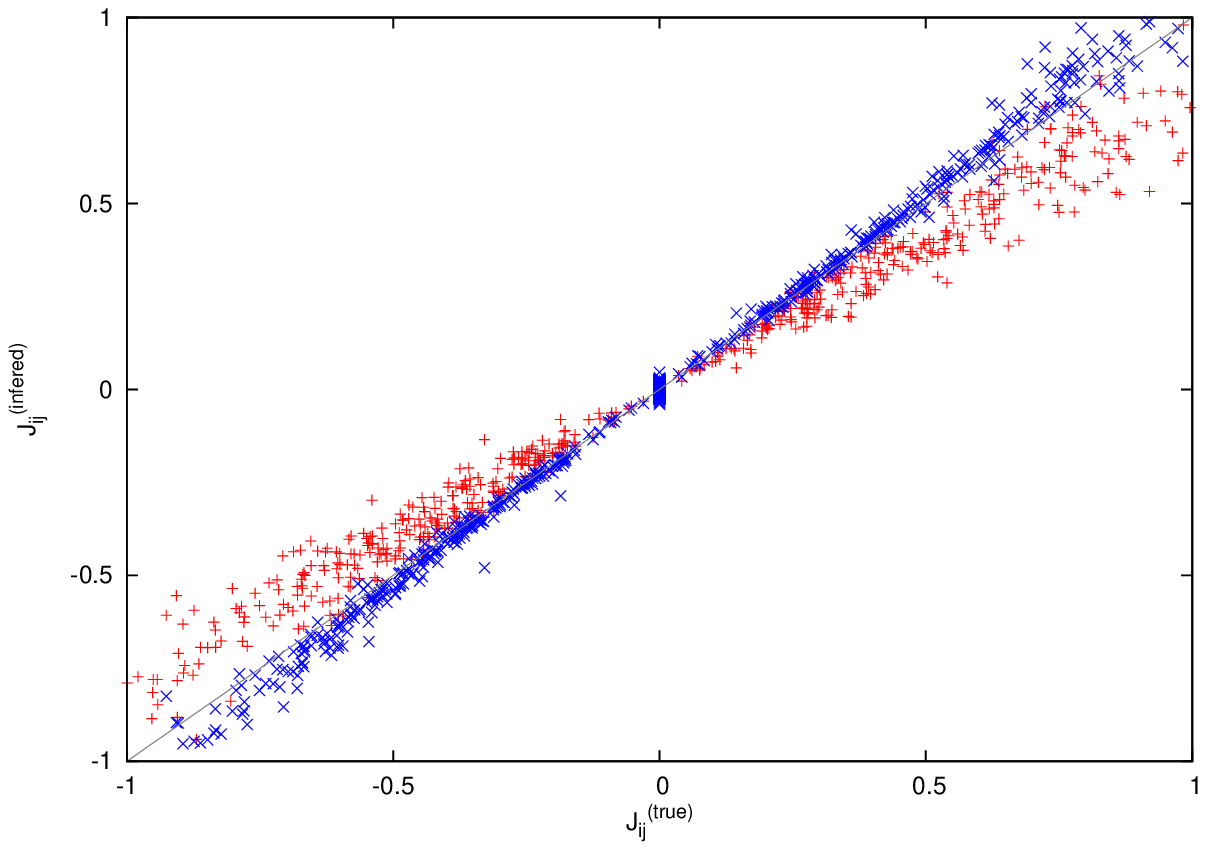}
  \caption{Mean field inference of a finite connectivity model. The
    couplings found by MF (blue) and nMF (red) are plotted versus the real
   couplings used to generate the data for a $N=200$  model on an asymmetric random regular graph
    with connectivity $c=6$ (average in-degree=$3$) , given $P=100000$ patterns generated at inverse temperature $\b=.6$. }
  \label{jjscat_sparse.eps}
\end{figure}
The method that we propose is exact and allows for a very precise
inference of the couplings when applied to the fully asymmetric SK spin glass,
at any temperature, if the number of patterns is large enough. At the
same time, it is an easy and versatile method which can be
used as an approximate inference method when the
number of patterns is not very  large (although one should at least have $P>N$ in
order for $C$ to be invertible), or when the underlying model is
not of the SK type. As an example showing the possible use of the
method, we have applied it to a sample where the $J_{ij}$ matrix is
sparse, generated as follows. We first generate a regular graph with $200$
vertices and degree $6$ on each vertex. For each edge $ij$ of this
graph we choose randomly with probability $1/2$ an orientation, say
$i\to j$. Then we take $J_{ji}=0$ and $J_{ij}$ is drawn randomly from
the probability density $(|x|/2)e^{-x^2/2}$. All the other couplings
corresponding to pairs of sites $kl$ which are not in the graph are
set to $0$. One then iterates the dynamics (\ref{Distr1Spin1Time}) 100000 times at
$\beta=.6$, and uses this data to reconstruct the couplings. Fig. 
\ref{jjscat_sparse.eps} shows the resulting couplings as a scatter
plot. The topology of the underlying interaction graph can be
reconstructed basically exactly, both by nMF and MF by using a threshold, deciding that
all reconstructed couplings with $|J_{ij}|<.04$ vanish. The non-zero
couplings are found accurately by the MF inference method.

To summarize, we have introduced a simple mean field method which can
be applied on a single instance of a dynamical fully asymmetric Ising
model. In the case of the asymmetric SK model this MF method gives the
exact values of the local magnetizations and the exact relation
between equal-time and time-delayed correlations. This method can be
used to solve efficiently the inverse problem, i.e. determine the
couplings and local fields from a set of patterns. Again this
inference method is exact in the limit of large sizes and large number
of patterns, in the asymmetric SK case. It can also be used in cases
where the underlying model is different, for instance for diluted
models. This could be quite useful for many applications. 

We thank Lenka Zdeborov\'a for useful discussions.This work has been supported in part by the EC grant 'STAMINA', No 265496.

\bibliographystyle{plain}
\bibliography{biblio}

\end{document}